\documentclass[twocolumn]{article}
\usepackage[a4paper, total={180mm, 243mm}]{geometry}
\usepackage[T1]{fontenc}
\usepackage{lipsum}
\usepackage{graphicx}
\usepackage{siunitx}
\usepackage{upgreek}
\usepackage{titling}
\usepackage{circledsteps}
\usepackage{subcaption}
\usepackage[version=4]{mhchem}
\usepackage{amsmath, amssymb}
\usepackage{soul}
\usepackage{empheq}

\definecolor{authorcolor}{cmyk}{0.8,0.6,0,0.3}
\newcommand{\eps}{\varepsilon}

\DeclareCaptionLabelSeparator{custom}{\textbar}
\DeclareCaptionFormat{custom}{\textsf{\textbf{#1 #2} \small #3}}
\graphicspath{{./figures}} 

\makeatletter
\newcommand{\showfontsize}{\f@size{} pt}
\makeatother

\usepackage[
    backend=biber,
    style=nature,
    url=false,
    isbn=false,
    date=year,
    maxnames=99
]{biblatex}
\let\cite\supercite

\newcommand{\reffig}[2]{Fig.~\ref{#1}{\textbf{#2}}}

\newcommand\identity{1\kern-0.25em\text{l}}

\title{Bright solitons in hybrid-dispersion photonic crystal microresonators}

\author{Alexander~E.~Ulanov$^{1}$,
        Lukas Bengel$^{2}$,
        Bastian Ruhnke$^{1}$,
        Alexandra Sakharova$^{1}$,
        Tobias Herr$^{1,3,*}$
}
\date{%
    \small $^1$Deutsches Elektronen-Synchrotron DESY, Notkestr. 85, 22607 Hamburg, Germany \\
    \small $^2$Institute for Analysis, Karlsruhe Institute of Technology KIT, Englerstrasse 2, 76131 Karlsruhe, Germany\\
    \small $^3$Institute of Experimental Physics, University of Hamburg UHH, Luruper Chaussee 149, 22761 Hamburg, Germany\\
    \small $^{*}$tobias.herr@desy.de
}

\begin{document}
\begin{refsection}

\maketitle

\textbf{
Bright dissipative Kerr solitons in optical microresonators provide chip-scale sources of ultrashort pulses and frequency combs. Their properties are defined by the cavity dispersion for which fundamentally conflicting requirements exist: short pulses and broadband spectra require weak dispersion, whereas strong dispersion is associated with predictable dynamics. Here, we resolve this conflict by introducing a localized strong-dispersion section spanning several modes around the pump resonance within an otherwise weakly dispersive system. We implement this hybrid-dispersion scheme in a photonic crystal microresonator and reveal a new soliton attractor of backward-propagating solitons, accessible at low pump power in a thermally stable manner within the blue-detuned regime. The conflicting requirements for broadband spectra and low-noise single-soliton formation are reconciled, even in microwave-repetition-rate resonators, which otherwise are prone to uncontrollable multi-soliton formation. These results highlight the potential to achieve previously incompatible characteristics in nonlinear photonic systems through hybrid-dispersion attractor shaping.}

\subsection*{Introduction}

Bright dissipative Kerr solitons (DKS) in continuous-wave (CW) driven optical microresonators enable low-timing jitter ultrashort pulse generation and broadband low-noise frequency combs\cite{kippenberg2018DissipativeKerrSolitons,pasquazi2018MicrocombsNovelGeneration, herr2026FrequencyCombsCoherent}. Small cavity size enables ultrahigh repetition rate above 10~GHz, which has enabled applications such as low-noise microwave generation\cite{liang2015HighSpectralPurity, sun2024IntegratedOpticalFrequency, kudelin2024PhotonicChipbasedLownoise, egbert2026AttosecondtimingMillimeterWaves}, optical communication\cite{marin-palomo2017MicroresonatorbasedSolitonsMassively} and computing\cite{xu202111TOPSPhotonic}, optical spectroscopy\cite{suh2016MicroresonatorSolitonDualcomb}, and astronomical spectrograph calibration\cite{obrzud2019MicrophotonicAstrocomb, suh2019SearchingExoplanetsUsing}.
The Kerr-nonlinear attractor state that underpins DKS formation is influenced by the resonator's mode dispersion, conveniently expressed by the integrated dispersion $D_\mathrm{int}(\mu) \approx \frac{1}{2}D_2\mu^2$, the deviations of the mode frequencies $\omega_\mu$ from a $D_1$-spaced frequency grid, where $D_1/(2\pi)$ is the resonator's free-spectral range (FSR), and $\mu$ the longitudinal mode number measured relative to the pumped mode. Positive values of $D_2$ correspond to anomalous group velocity dispersion that is needed to support bright soliton pulses.
When designing microresonators for DKS generation, a fundamental conflict exists: weak dispersion $0<D_2\ll\kappa$ ($\kappa$ is the resonance width) is needed to support ultrashort pulses and broadband spectra, but strong dispersion $D_2\sim\kappa$ is needed to create predictable initiation of low-noise oscillation\cite{herr2012UniversalFormationDynamics} and single-pulse formation\cite{yu2021SpontaneousPulseFormation, ulanov2024SyntheticReflectionSelfinjectionlocked}.

Recently, new opportunities for dispersion engineering have emerged by exploiting mode hybridization and associated mode frequency shifts (mode splitting/avoided mode crossing). Approaches include deliberate coupling between transverse mode families\cite{yang2016SpatialmodeinteractioninducedDispersiveWaves, liu2025BreakingBandwidthefficiencyTradeoff, zheng2026EngineeredModeCoupling}, mode coupling between resonators\cite{helgason2021DissipativeSolitonsPhotonic, liu2025NearvisibleIntegratedSoliton} (photonic molecules), or photonic crystal resonators (PhCRs), where a waveguide corrugation induces controllable coupling between one or multiple selected forward and backward propagating modes\cite{yu2021SpontaneousPulseFormation, black:2022,lucas2023TailoringMicrocombsInversedesigned}. Specifically, if the pump mode is shifted to lower frequencies, this can drastically alter the system's characteristics: Noisy states of the system can be suppressed\cite{yu2021SpontaneousPulseFormation} and effectively blue-detuned operation, which overcomes thermally unstable DKS initiation, can be achieved\cite{nishimoto2025SelfcoolingBluedetunedDissipative}. However, these functionalities come at the expense of drastically increased threshold power, as differential self- and cross-phase modulation must compensate the frequency shift of the pump resonance. As a frequency-shifted pump mode implies strong dispersion at the pump wavelength, the first oscillating sidebands can be moved closer to the pump\cite{herr2012UniversalFormationDynamics}. This may enforce deterministic generation of well-defined single-DKS\cite{yu2021SpontaneousPulseFormation,ulanov2024SyntheticReflectionSelfinjectionlocked}. However, native $D_2$ must not be too small to ensure mode-selectivity for the first oscillating sidebands. This is again in conflict with a low-$D_2$ for broadband operation and is incompatible with technologically important microwave repetition rate systems, where $D_2$ is intrinsically low.

\begin{figure*}[!t]
  \centering
  \includegraphics[width=\textwidth]{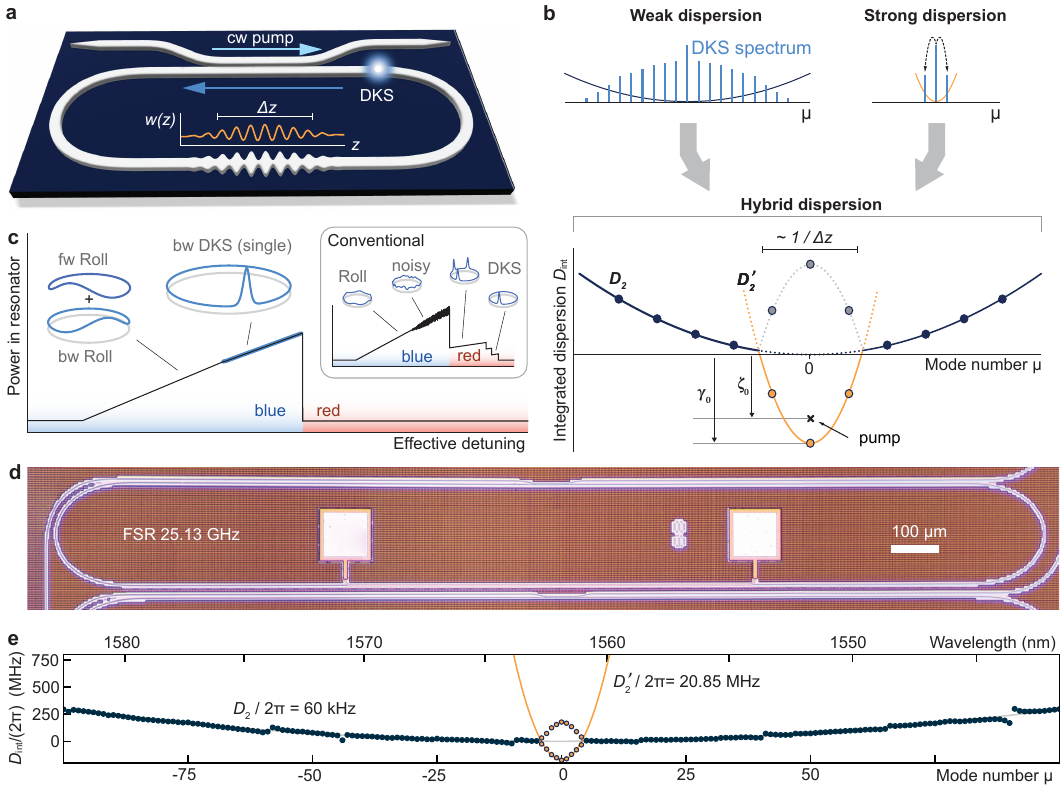}
  \caption{\textbf{Hybrid dispersion in a photonic crystal microresonator.}
    \textbf{a}, Illustration of a photonic crystal microresonator with an amplitude-modulated waveguide corrugation pattern creating a hybrid dispersion profile. The waveguide corrugation with amplitude $w(z)$ extends over a (full-width half maximum) distance $\Delta z$, where $z$ is a path length parameter of the resonators. When pumped by a forward (FWD) propagating pump laser, the hybrid dispersion enables blue-detuned backward (BWD) propagating solitons.
    \textbf{b}, Hybrid dispersion combines weak and strong dispersion characteristics in a single resonator, associated with broadband and narrowband optical spectra.
    It is defined by a weak native dispersion $D_2$, a pump mode frequency shift $\gamma_0$, and a strong induced dispersion $D_2'$ around the central pump mode; the width of the strong dispersion interval scales inversely with $\Delta z$. A blue-detuned pumping situation with detuning $\zeta_0<\gamma_0$ is indicated.
    \textbf{c}, Hybrid dispersion leads to distinct DKS formation dynamics involving hybrid FWD and BWD rolls (Turing patterns), and DKS formation propagating in the BWD direction, all within the blue-detuned pumping regime. This is in stark contrast to the conventional dynamics involving rolls, noisy and stochastic multi-DKS states (inset). 
    \textbf{d}, Photograph of a silicon-nitride hybrid-dispersion microresonator with a free-spectral range (FSR) of 25.13\,GHz.
    \textbf{e}, Measured integrated dispersion $D_\mathrm{int}$ of the resonator shown in d. Each dot marks a resonance (dark blue: native dispersion $D_2$; orange: modified resonance frequencies). Gray and orange lines are a guide to the eye, highlighting the integrated dispersion parabolas. 
    }
    \label{fig:concept}
\end{figure*}

Here, we introduce the concept of hybrid dispersion: globally weak dispersion combined with strong dispersion in a narrow spectral window around the pump. This approach reconciles previously competing dispersion requirements within a single microresonator.  
In a combined numerical, analytical, and experimental approach, we show that this dispersion landscape gives rise to a new attractor of backward-propagating DKS, enabling thermally stable, blue-detuned single-soliton operation at low pump power
without the canonical ``soliton step'' of conventional red-detuned DKS. Crucially, this behavior is achieved in low-$D_2$ microwave repetition rate (25~GHz) resonators, which are otherwise prone to random multi-DKS formation. 
Our results address key challenges in DKS sources and point to hybrid-dispersion attractor shaping as a route to nonlinear photonic systems with previously incompatible characteristics.

\begin{figure*}[!t]
  \centering
  \includegraphics[width=\textwidth]{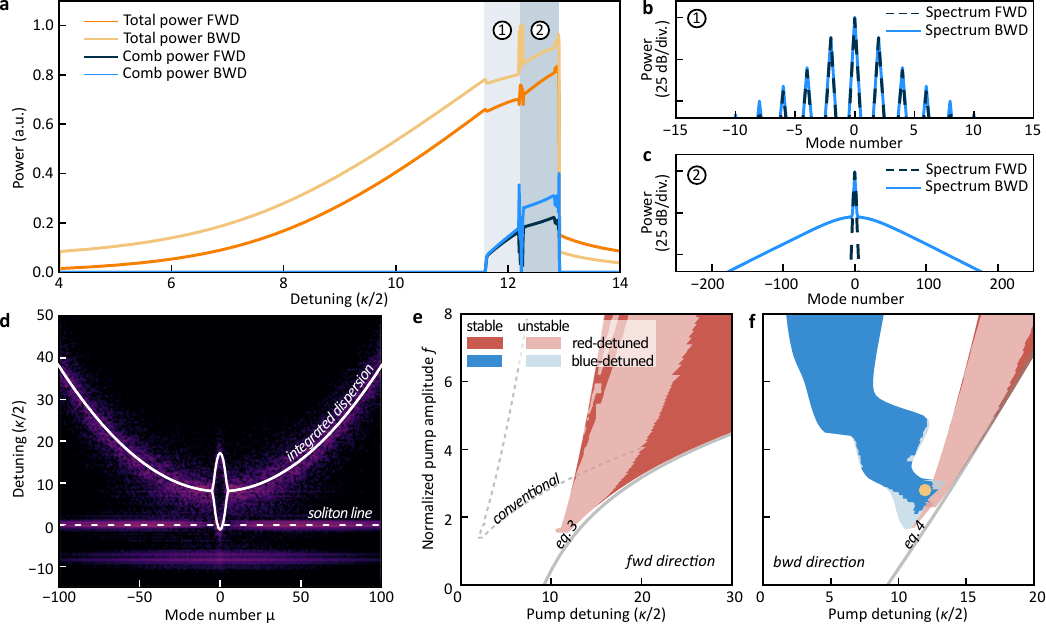}
    \caption{\textbf{Numerical simulations.}
      \textbf{a}, Total intracavity power in the forward (FWD, orange) and backward (BWD, yellow) directions, and the corresponding Kerr frequency comb powers (intracavity power in the FWD and BWD directions excluding the pump line, dark blue and light blue), recorded from numerical integration of a system of 1024 coupled-mode equations during a pump-laser frequency scan across the lower-frequency hybrid mode.
      \textbf{b}, \textbf{c}, Optical spectra in the FWD and BWD directions (dashed and solid lines, respectively) for states \Circled{1} and \Circled{2} from panel a.
      \textbf{d}, Numerically reconstructed nonlinear dispersion relation for comb state \Circled{2}, with the effective integrated dispersion and soliton line highlighted by solid and dashed white lines, respectively.
      \textbf{e}, \textbf{f}, Soliton existence ranges in the FWD and BWD directions, computed using bifurcation analysis. Blue and red areas correspond to solutions with effective blue and red detuning relative to the pumped resonance. Darker areas indicate stable solutions, while lighter areas indicate solutions that are unstable due to oscillations. Solid gray lines show the analytically derived existence boundaries (see Methods). The yellow dot marks the state \Circled{2} from panel c. The range enclosed by the gray dashed lines (in panel e) indicates the conventional single-soliton existence range (no resonance hybridization).
      }
    \label{fig:theory}
\end{figure*}

\begin{figure*}[!t]
  \centering
  \includegraphics[width=\textwidth]{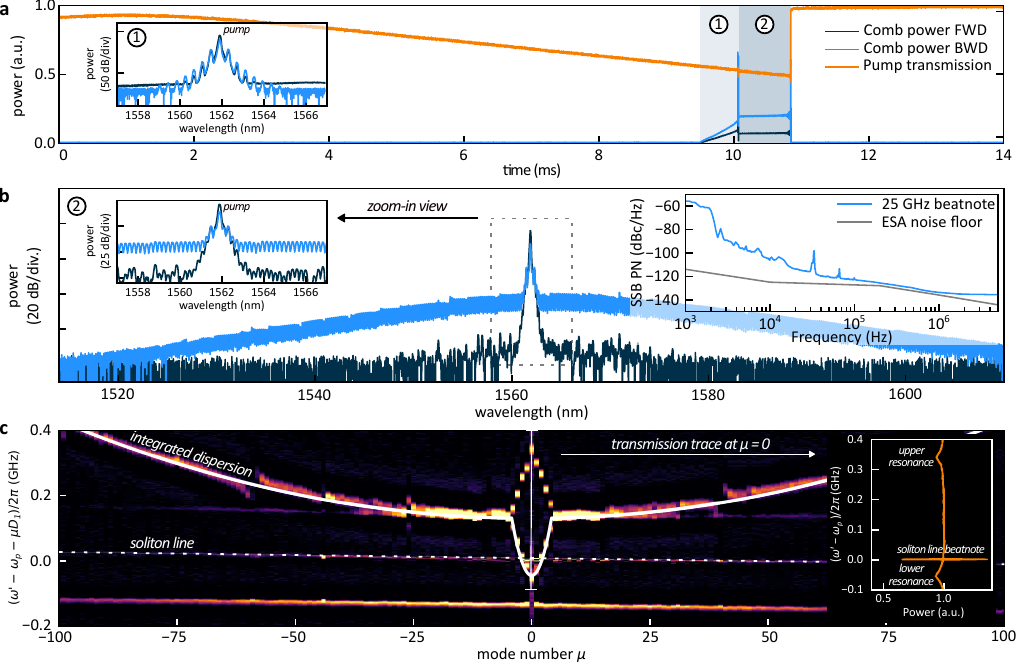}
  \caption{\textbf{Experimental results.}
    \textbf{a}, Transmission (orange) and generated comb powers in the forward (FWD, dark blue) and backward (BWD, light blue) directions, recorded during a pump-laser frequency scan across the central mode of the steep-dispersion section from blue to red detuning. The left inset shows the recorded optical spectra in the FWD (dark blue) and BWD (light blue) directions for comb state \Circled{1}.
    \textbf{b}, Optical spectra in the FWD (dark blue) and BWD (light blue) directions for comb state \Circled{2} from panel a. The right inset shows the single-sideband phase noise of the 25~GHz repetition-rate beatnote, measured on a fast photodiode and recorded using an electrical spectrum analyzer (ESA). The gray line indicates the ESA measurement floor.
    \textbf{c}, Experimentally reconstructed nonlinear dispersion relation, obtained by scanning an auxiliary continuously tunable probe laser across the Kerr comb lines in state \Circled{2} from panel a and recording its transmission on a slow photodetector. The effective integrated dispersion and the DKS frequencies (``soliton line'') are highlighted by solid and dashed white lines, respectively. The inset shows the auxiliary-laser transmission in the vicinity of the pumped resonance at $\mu=0$.
  }
    \label{fig:exp}
\end{figure*}

\subsection*{Results}
To implement hybrid dispersion, we utilize racetrack PhCRs with 25~GHz FSR. The microresonator has a corrugation with a spatial period equal to half the pump-resonance wavelength, and with a smooth pulse-like amplitude envelope (\reffig{fig:concept}{a}). Similar to segmented corrugation\cite{li2025BroadbandAccurateElectric, kheyri2025OnchipMicroresonatorDispersion}, or the superposition of multiple corrugation periods~\cite{lucas2026InducedDirectionalSwitching}, this effectively confines the corrugation to part of one straight waveguide section. Decomposing the corrugation into spatial Fourier components, each component couples a pair of frequency-degenerate forward- (FWD) and backward-propagating (BWD) modes, hybridizing them and splitting the corresponding resonances. The magnitude of each Fourier component sets the FWD-BWD coupling rate $\gamma_\mu$, and hence the resonance frequency shift, allowing tailored dispersion landscapes\cite{lucas2023TailoringMicrocombsInversedesigned}. 

We leverage this to combine globally weak dispersion $D_\mathrm{int}(\mu)=\frac{1}{2}D_2\mu^2$ with a strong localized dispersion $D_\mathrm{int}'(\mu)=\frac{1}{2}D_2'\mu^2-\gamma_0$ in a narrow spectral window around the pump, where $D_2'\gg D_2$ and $\gamma_0$ is the coupling rate between FWD and BWD fields at $\mu=0$; in our case, the pulse like corrugation implies that all spatial Fourier components (and hence the FWD-BWD coupling) are in phase. Effectively this results in two integrated-dispersion parabolas offset by $\gamma_0$ (\reffig{fig:concept}{b}). In a minimal implementation, the corrugation pattern requires only three Fourier components, through which $D_2'$ and $\gamma_0$ can be set independently. As we show below, this dispersion landscape enables blue-detuned, BWD DKS generation, seeded by narrow-band FWD-BWD Turing rolls (\reffig{fig:concept}{c}).

The resonators are fabricated in a commercial wafer-level low-loss silicon nitride (Si$_3$N$_4$) integrated photonic platform (\reffig{fig:concept}{d}).
The waveguide cross-section is $2\times0.8~\mu m^2$ and Euler bends minimize mixing of higher-order transverse modes. The resonance width is $\kappa/2\pi \approx 40$~MHz (critically coupled), corresponding to a loaded Q-factor of $\approx 5$~million, and the native resonator dispersion is $D_2/2\pi \approx 60$~kHz. The spatial period of the corrugation is 426~nm, and its pre-lithography design envelope is described by $\Delta w \sin^2(\pi z/z_0)$, where $\Delta w$ ranges from 10 to 100\,nm, and $z_0 = 1065~\mu$m is the total grating length.
A set of 10 resonators is fabricated where we sweep the amplitude of the corrugation pattern $\Delta w$ to scale both $\gamma_0$ and $D_2'$. To characterize the resulting hybrid dispersion profile, we perform frequency-comb-assisted laser frequency scans with a continuously tunable laser \cite{delhaye:2009} as shown in \reffig{fig:concept}{e} for one example with a splitting (coupling rate) of $\gamma_0/2\pi\approx 175$~MHz. Noticeable splitting extends across 9 modes in the vicinity of 1562~nm, producing an effective steep dispersion section with $D'_2/2\pi \approx 20$~MHz, more than two orders of magnitude larger than the native dispersion and comparable with the cavity linewidth. Besides deliberate mode hybridization, the dispersion landscape is nearly free of avoided mode crossings (AMXs). In the fabricated resonators 
$\gamma_0$ approximately ranges from 50 to 600~MHz, and $D_2'$ from 6 to 80~MHz.

To guide our experiments, and following the canonical procedure for DKS generation, we numerically simulate the intracavity dynamics under CW driving. The pump frequency is scanned across the lower-frequency hybrid mode, from blue- toward red-detuning (i.e. increasing $\zeta_0$). The simulation is based on the coupled-mode equations, takes FWD and BWD fields into account, and uses the parameters of the fabricated PhCRs (see Methods for model details and simulation parameters). For weak hybrid dispersion profiles, we observe the canonical dynamics: noisy states and stochastic multi-DKS formation~\cite{herr2014TemporalSolitonsOptical}. 
%single-FSR spaced sidebands or Turing patterns, but no broadband DKS in the latter. In contrast, for intermediate hybrid dispersion,
In contrast, for pronounced hybrid dispersion we observe a qualitatively new behavior. \reffig{fig:theory}{a} shows the detuning-dependent total intracavity FWD and BWD powers, as well as the respective FWD and BWD comb powers (excluding the pump). As the detuning increases and approaches the effective resonance frequency of the pumped hybrid mode, the intracavity power increases in both FWD and BWD directions, with slightly higher power in the BWD direction. At a threshold detuning in the blue-detuned regime, a primary sideband Turing pattern emerges in both FWD and BWD directions, including modes $\mu = \pm 2, \pm4, ..$ (state \Circled{1} in \reffig{fig:theory}{a, b}). As the detuning is increased further, the system transitions directly into a single-soliton state propagating in the BWD direction (state \Circled{2} in \reffig{fig:theory}{a, c}), while a fundamental narrowband Turing roll pattern exists in the FWD direction. Notably, in the transition to the DKS state, the abrupt drop in intracavity power (``soliton step''\cite{herr2014TemporalSolitonsOptical}) characteristic of conventional DKS generation is absent. This \emph{step-free} soliton initiation is strong indication of effectively blue-detuned pumping, a regime that is of high interest for its thermal stability and self-cooled low-noise operation\cite{nishimoto2025SelfcoolingBluedetunedDissipative}. 

To corroborate the hypothesis of blue-detuned pumping, we compute the nonlinear dispersion relation\cite{leisman2019EffectiveDispersionFocusing,anderson2023DissipativeSolitonsSwitching,herr2026FrequencyCombsCoherent} numerically for the comb state \Circled{2}. The results presented in \reffig{fig:theory}{d} show the soliton line (dashed white) relative to the (effective) integrated dispersion (solid white). The soliton line, which includes the pump laser at $\mu = 0$, lies above the lower frequency branch of the integrated dispersion, explicitly confirming the blue-detuned pumping in the simulation. Overall, the dynamics can be understood as follows: The initial sideband formation is governed by the steep dispersion $D_2'$ around the pump laser, which forces sideband initiation close to it --- a key criterion for low noise in non-DKS states~\cite{herr2012UniversalFormationDynamics}. As the detuning increases further, the parameter range of a BWD soliton attractor is reached, leading to a transition from the Turing pattern to a single BWD DKS.

To understand how the hybrid dispersion profile reshapes the soliton attractors, we perform a bifurcation analysis (see Methods and Supplementary Information). We identify the values of pump laser detuning and power for which stable DKS solutions exist. By considering the spectral mode fields of these solutions and computing the corresponding pumped-resonance Kerr frequency shifts, we also determine whether the pump laser is effectively blue or red detuned. 
Consistent with previous work\cite{helgason2023SurpassingNonlinearConversion}, we find red-detuned soliton solutions in the FWD direction; their existence range is shifted towards larger detuning due to frequency shift of the pump resonance (\reffig{fig:theory}{e}). 
Strikingly, the bifurcation analysis also reveals the existence of a new blue-detuned soliton attractor of BWD-propagating DKS (\reffig{fig:theory}{f}), confirming the result of the dynamic simulation. Additionally, we derive analytic existence boundaries for FWD and BWD DKS, which are in excellent agreement with the numeric bifurcation analysis. The analytic boundaries are derived in the long cavity limit for large detuning and pump strength (\reffig{fig:theory}{e, f}; see Methods and Supplementary information, SI).

Next, we proceed with experimental generation of DKS in the hybrid-dispersion resonator, that corresponds to the dispersion profile in \reffig{fig:concept}{e}. We slowly scan a tunable CW laser across the central mode of the steep dispersion section from blue to red detuning with an on-chip pump power of only 60~mW. The large pump-mode splitting enables us to avoid thermal shadowing effects \cite{yu2021SpontaneousPulseFormation} reported previously and to scan the laser only across the lower branch of the split Lorentzian lineshape, greatly simplifying the nonlinear dynamics and practical operation of such devices. During the scan, we simultaneously monitor the transmission and the generated comb power in both the FWD (co-propagating with the pump) and BWD directions. The resulting traces are shown in \reffig{fig:exp}{a}. Closely resembling the numeric simulation (\reffig{fig:theory}{a}), there are two characteristic regions corresponding to distinct comb states (\Circled{1} and \Circled{2}). State \Circled{1} (\reffig{fig:exp}{a}, left inset) corresponds to a two-roll (Turing pattern) comb state observed in both the FWD and BWD directions. State \Circled{2} (\reffig{fig:exp}{b}) corresponds to a single-soliton state in the BWD direction with on-chip comb power (excluding the pump line) of $\approx 2$~mW and its backscattered replica in the FWD direction. Crucially, and consistent with the simulation, an abrupt soliton-step upon DKS formation is absent in the transmission trace, in sharp contrast to conventional bright DKS generation. Thanks to this ``step-free generation'', it is possible to scan the laser arbitrarily slowly into the resonance and simply halt the laser frequency scan at any position along the trace, greatly simplifying DKS generation. 

Moreover, blue-detuned pumping is associated with self-cooling of the resonator which can lower the repetition rate phase noise\cite{nishimoto2025SelfcoolingBluedetunedDissipative, carmon2004DynamicalThermalBehavior}. To characterize this noise, we amplify the generated spectrum of state \Circled{2} with an EDFA (after rejecting the pump line) and measure the repetition-rate beatnote with a fast photodetector. We record the corresponding IQ data using an ESA and obtain a phase-noise power spectral density of -110\,dBc/Hz at 10 kHz and below -120\, dBc/Hz at 100kHz, confirming low-noise operation (\reffig{fig:exp}{b}, right inset). 

Finally, to experimentally confirm the effective blue detuning with respect to the pump resonance, we measure the nonlinear dispersion relation by sweeping the frequency $\omega'$ of a second continuously tunable probe laser (injected via a circulator in the direction counter-propagating to the main pump), and monitoring its transmission on a photodetector \cite{wildi2023SidebandInjectionLocking}. The data is presented in \reffig{fig:exp}{c}, clearly showing the integrated dispersion corresponding to \reffig{fig:concept}{e} and the soliton line. The inset shows the probe-laser transmission in the vicinity of the pumped resonance, $\mu = 0$. Both upper and lower sidebands of the split Lorentzian lineshape are clearly resolved, along with the beatnote between the probe laser and the soliton (pump) line. These observations provide direct evidence that the pump laser operates at an effective blue detuning relative to the hot-cavity resonance frequency.

\subsection*{Conclusion}

We have introduced and demonstrated hybrid dispersion in a photonic crystal microresonator, where weak global dispersion is combined with strong localized dispersion around the pump, and shown that it gives rise to an attractor of backward-propagating dissipative Kerr solitons. 
The large frequency-shift value $\gamma_0$ of the pump mode enables fully blue-detuned DKS, a regime that is thermally stable and where a canonical soliton-step upon DKS generation is absent. The local dispersion described by $D_2'$, ensures that this regime can be accessed easily and at low pump power. Moreover, the large value of $D_2'\gg D_2$ forces the system to operate in the desirable single DKS state. 
We demonstrate this in a microwave-repetition-rate (25~GHz) resonator whose intrinsically weak dispersion otherwise leads to random multi-soliton formation. While crucial for DKS initiation, $D_2'$ has little influence on the final broadband DKS state, which is largely defined by the system parameters $D_2$, $\gamma_0$, and the operating parameters $\zeta_0$ and $f$. As such the parameters $D_2'$ and $\gamma_0$ effectively decouple soliton initiation from the final soliton state. This is consistent with our analysis in the long cavity limit, which is valid for DKS spectra whose bandwidth significantly exceeds the modified dispersion interval around the pump laser (see SI). The ability to independently control $D_2'$ and $\gamma_0$ opens a powerful degree of freedom for next-generation chip-integrated pulse and frequency comb sources. More broadly, our results point towards hybrid-dispersion as a powerful approach to create new attractor states in nonlinear systems that combine previously incompatible characteristics.
% locally structured dispersion as a means to shape the soliton attractor itself, pointing to hybrid-dispersion attractor shaping as a route to nonlinear photonic systems that combine previously incompatible characteristics.

\subsection*{Methods}
\small

\paragraph{System of coupled mode equations.} 
To simulate the nonlinear dynamics shown in \reffig{fig:theory}{a}, we numerically integrate a system of coupled-mode equations describing 512 forward- (FWD) and 512 backward-propagating (BWD) mode amplitudes $a_\mu$ and $b_\mu$, respectively, where $\mu$ is the relative (longitudinal) mode number measured with respect to the pump mode ($\mu = 0$):
\begin{align}\label{eq:CoupledMode}
\begin{split}
  \partial_t a_\mu = &-(1+\mathrm{i}\zeta_\mu)a_\mu + \mathrm{i}\sum_{\nu,\eta} a_\nu a_\eta a_{\nu+\eta-\mu}^* + 2\mathrm{i} a_\mu \sum_{\eta} |b_\eta|^2 \\
  &+ \mathrm{i}\beta_\mu b_\mu + f \delta_{\mu 0}, \\
  \partial_t b_\mu = &-(1+\mathrm{i}\zeta_\mu)b_\mu + \mathrm{i}\sum_{\nu,\eta} b_\nu b_\eta b_{\nu+\eta-\mu}^* + 2\mathrm{i} b_\mu \sum_{\eta} |a_\eta|^2 \\
  &+ \mathrm{i} \beta_\mu a_\mu,
\end{split}
\end{align}
where $\zeta_\mu = \frac{2}{\kappa} (\omega_{\mu} - \omega_p - \mu D_1)$ is the dimensionless detuning defined by the pump laser frequency $\omega_p$ and the resonance frequencies $\omega_\mu$; $f = \sqrt{8\eta\,\omega_0 c n_2 P/(\kappa^2 n^2 V_\mathrm{eff})}$ is the normalized pump power, where $\eta = 1/2$ is the coupling coefficient (critical coupling), $c$ the speed of light, $P$ the pump power, $n$ the refractive index, $n_2$ the nonlinear refractive index, and $V_\mathrm{eff}$ the effective mode volume. The third term on the right-hand side of each equation corresponds to the cross-phase modulation by the counter-propagating waves. The fourth term describes linear coupling between the FWD and BWD waves, with $\beta_\mu = 2\gamma_\mu/\kappa$ the dimensionless FWD-BWD coupling rate ($\gamma_\mu$ being the mode-dependent backscattering rate); we assume that $\beta_\mu \in \mathbb{R}$. 

Simulation parameters are set similar to those of the experimental system, including $d_2 = 2D_2/\kappa = 0.006$ and $f = 2.8$. Within the mode spectrum, 9 FWD-BWD mode pairs are coupled, forming the locally steep dispersion section with $d'_2 = 2D'_2/\kappa = 0.9$ and $\beta_0 = 9.2$. The numerical simulation also allows us to compute the nonlinear dispersion relation as illustrated in \reffig{fig:theory}{d}.

\paragraph{Numerical bifurcation analysis.}
We systematically compute the DKS existence ranges (see~\reffig{fig:theory}{e, f}) using numerical bifurcation analysis. For the implementation, it is advantageous to consider the time-domain formulation of~\eqref{eq:CoupledMode}, given by
\begin{align}\label{eq:CoupledLLE}
\begin{split}
  \partial_t a = &-(1+\mathrm{i}\zeta_0 - \mathrm{i} d_2 \partial_\tau^2)a + \mathrm{i}\big(|a|^2 + \frac{2}{2\pi} \int_{-\pi}^{\pi} |b(t,\tau')|^2 \mathrm{d}\tau' \big) a \\
  &+ \mathrm{i} \frac{1}{2\pi} \sum_\mu \beta_\mu\int_{-\pi}^{\pi} b(t,\tau') \mathrm{e}^{\mathrm{i}\mu (\tau-\tau')} \mathrm{d} \tau' + f , \\
  \partial_t b = &-(1+\mathrm{i}\zeta_0 - \mathrm{i} d_2 \partial_\tau^2)b + \mathrm{i}\big(|b|^2 + \frac{2}{2\pi}\int_{-\pi}^{\pi} |a(t,\tau')|^2 \mathrm{d}\tau' \big) b \\
  &+ \mathrm{i} \frac{1}{2\pi} \sum_\mu \beta_\mu\int_{-\pi}^{\pi} a(t,\tau') \mathrm{e}^{\mathrm{i}\mu (\tau-\tau')} \mathrm{d}\tau' ,
\end{split}
\end{align}
for the FWD and BWD fields $a(t,\tau) = \sum_{\mu} a_\mu(t) \mathrm{e}^{\mathrm{i} \mu \tau}$ and $b(t,\tau) = \sum_{\mu} b_\mu(t) \mathrm{e}^{\mathrm{i} \mu \tau}$. Since both integrated dispersion profiles $D_\text{int}(\mu)$ and $D_\text{int}'(\mu)$ are symmetric in $\mu$, we find $\beta_\mu = \beta_{-\mu}$ and thus the coupled system~\eqref{eq:CoupledLLE} is reversible in $\tau$. As a consequence, DKS correspond to stationary solutions obtained by solving~\eqref{eq:CoupledLLE} with $\partial_t a = \partial_t b = 0$. 
Numerically, we discretize the stationary equation in $\tau$ using linear finite elements on a uniform mesh with 512 grid points. All nonlocal integral terms are approximated using the trapezoidal rule. To determine the DKS existence ranges, we design a three-step continuation algorithm, which is based on the MATLAB bifurcation packages pde2path~\cite{uecker2014Pde2pathMatlabPackage}.

In Step~1, we compute an initial solution for the continuation, either by time-integrating~\eqref{eq:CoupledMode} until a DKS state is reached or by using the DKS approximation formulas derived in the SI. Starting from this initial seed, we compute in  Step~2 a solution branch reaching from the minimal possible value $f_\textup{min}$ to a prescribed maximal value $f_\textup{max}$ by an alternating continuation in $\zeta_0$ and $f$. For each solution obtained in Step 2, we perform in Step 3 continuation in $\zeta_0$ in both positive and negative directions until a fold bifurcation is reached. The union of all solutions computed in Step 3 defines the DKS existence region. Note that for different initial seeds, different existence regions may arise which are not necessarily connected by a solution branch. For our simulations, we take a nonuniform discretization of the $\zeta_0$-$f$-plane. In the $f$-direction, we use grid spacings of $0.05$ for $f \leq 4$, $0.1$ for $4 \leq f \leq 6$, and $0.2$ for $f \geq 6$. The $\zeta_0$-direction is discretized by the path-continuation step size of $0.2$. The parameters $d_2,d_2'$ and $\beta_0\geq0$ are set to the same values as for the time integration simulations.

\paragraph{Boundaries of the DKS existence ranges.}
Under the standard assumption $d_2,\zeta_0>0$ for bright solitons~\cite{herr2014TemporalSolitonsOptical}, we obtain fully explicit formulas for the existence boundaries of FWD- and BWD-propagating DKS. The derivations are based on perturbation theory and require the nondegeneracy condition $\zeta_0^2 \neq \beta_0^2$ (see the SI for the technical details).
For FWD-propagating DKS this yields the existence condition
\begin{align}\label{eq:DKS_condition_forward}
    2^{3/2}\frac{|\zeta_0^2 - \beta_0^2|}{\zeta_0^{3/2}} \leq f \pi.
\end{align}
Notably, only the central coupling coefficient $\beta_0$ enters~\eqref{eq:DKS_condition_forward}, indicating that hybridization of the pumped mode provides the dominant contribution to the reshaping of the DKS existence region. For $\beta_0 = 0$, the condition~\eqref{eq:DKS_condition_forward} reduces to $2^{3/2} \zeta_0^{1/2} \leq f\pi$, consistent with previous findings for conventional resonators~\cite{herr2014TemporalSolitonsOptical,bengel2025ExistenceStabilitySolitonbased}. For increasing values of $\beta_0$, inequality~\eqref{eq:DKS_condition_forward} predicts a shift of the DKS existence region towards larger detuning values, in agreement with the simulation in \reffig{fig:theory}{e}. 

For BWD-propagating DKS, the existence condition reads
\begin{align}\label{eq:DKS_condition_backward}
2^{3/2}\frac{|\zeta_0^2 - \beta_0^2|}{\sqrt{\zeta_0}} \leq \beta_0 f \pi.
\end{align}
As in the FWD case, only the central coupling coefficient $\beta_0$ enters the formula. Moreover, the BWD DKS existence region shifts towards larger detuning values as $\beta_0$ increases, whereas BWD-propagating DKS disappear in the limit where $\beta_0$ approaches zero. Finally, the perturbation analysis in the SI shows that both conditions~\eqref{eq:DKS_condition_forward} and~\eqref{eq:DKS_condition_backward} provide asymptotically sharp descriptions of the existence boundaries in the limit of large detuning and pump strength, while remaining in excellent agreement with numerical simulations already for small and moderate parameter values.

% \subsection*{Author Contributions}
% A.U. performed the experiments and analysed the data. L.B. performed the numeric and the analytic analysis. B.R. and A.S. designed the photonic chip. T.H. supervised the project. A.U., L.B., and T.H. wrote the manuscript with input from all authors.

\small
\paragraph{Funding.}
\small
This project has received funding from the European Research Council (ERC) under the EU's Horizon 2020 research and innovation program (grant agreement No 853564), from the EU's Horizon 2020 research and innovation program (grant agreement No 101137000), from the Deutsche Forschungsgemeinschaft (DFG, German Research Foundation) – Project-ID 258734477 – SFB 1173, and through the Helmholtz Young Investigators Group VH-NG-1404; the work was supported through the Maxwell computational resources operated at DESY.

% \subsection*{Disclosures}
% \small All authors declare no conflict of interest.

\printbibliography

\end{refsection}

\newpage
\onecolumn
\begin{refsection} 
\setcounter{equation}{0}
\setcounter{figure}{0}
% Figure numbering style with "S"
\renewcommand{\thefigure}{S\arabic{figure}}

% \documentclass[onecolumn]{article}
% \usepackage[a4paper, total={180mm, 243mm}]{geometry}
% \usepackage[T1]{fontenc}
% \usepackage{lipsum}
% \usepackage{graphicx}
% \usepackage{siunitx}
% \usepackage{upgreek}
% \usepackage{titling}
% \usepackage{circledsteps}
% \usepackage{subcaption}
% \usepackage[version=4]{mhchem}
% \usepackage{amsmath, amssymb}
% \usepackage{soul}
% \usepackage{empheq}

% \definecolor{authorcolor}{cmyk}{0.8,0.6,0,0.3}
% \newcommand{\eps}{\varepsilon}

% % Line numbering
% \usepackage[switch]{lineno} 
% \linenumbers
% \modulolinenumbers[5]
% \renewcommand\linenumberfont{\normalfont\scriptsize\sffamily}

% % Tweaking Captions
% \DeclareCaptionLabelSeparator{custom}{\textbar}
% \DeclareCaptionFormat{custom}{\textsf{\textbf{#1 #2} \small #3}}
% \captionsetup{format=custom, labelsep=custom}

% % Required for the .pdf_tex figures to work
% \graphicspath{{./figures}} 

% % To be able to see the current font size
% \makeatletter
% \newcommand{\showfontsize}{\f@size{} pt}
% \makeatother

% % BibLaTeX.
% \usepackage[
%     backend=biber,
%     style=nature,
%     url=false,
%     isbn=false,
%     date=year,
%     maxnames=99
% ]{biblatex}
% \let\cite\supercite

% \addbibresource{all_mod.bib}

% % For citing subfig
% \newcommand{\reffig}[2]{Fig.~\ref{#1}{\textbf{#2}}}
% \newcommand{\bhat}[1]{\hat{\textbf{\textrm{#1}}}}
% \newcommand\identity{1\kern-0.25em\text{l}}

\title{Bright solitons in hybrid-dispersion photonic crystal microresonators - \\ Supplemental Information}

% \author{Alexander~E.~Ulanov$^{1}$,
%         Lukas Bengel$^{2}$,
%         Bastian Ruhnke$^{1}$,
%         Alexandra Sakharova$^{1}$,
%         Tobias Herr$^{1,3,*}$
% }
% \date{%
%     \small $^1$Deutsches Elektronen-Synchrotron DESY, Notkestr. 85, 22607 Hamburg, Germany \\
%     \small $^2$Institute for Analysis, Karlsruhe Institute of Technology KIT, Englerstrasse 2, 76131 Karlsruhe, Germany\\
%     \small $^3$Institute of Experimental Physics, University of Hamburg UHH, Luruper Chaussee 149, 22761 Hamburg, Germany\\
%     \small $^{*}$tobias.herr@desy.de
% }

% \begin{document}

\maketitle

\section{Analytic description of FWD- and BWD-propagating DKS and derivation of DKS existence boundaries}

Here, we derive the analytic formulas for the soliton existence ranges (inequalities (3) and (4) in the Methods Section) and obtain approximation formulas for FWD and BWD solitons. 
We remark that semi-analytic expressions for soliton existence ranges have recently been derived in the special case where only the pump mode is hybridized~\cite{sakaue2026CoupledLugiatoLefever}.
Let us start from the stationary coupled Lugiato-Lefever system given by
\begin{align}\label{eq:LLE_Stationary}
\begin{split}
  \mathrm{i} d_2 \partial_\tau^2 a -(\eps+\mathrm{i}\zeta_0 )a + \mathrm{i}\big(|a|^2 + \frac{2}{2\ell} \int_{-\ell}^{\ell} |b(\tau')|^2 \mathrm{d}\tau' \big) a
  + \mathrm{i} \frac{1}{2\ell} \sum_\mu \beta_\mu\int_{-\ell}^{\ell} b(\tau') \mathrm{e}^{\mathrm{i}\mu \frac{\pi}{\ell} (\tau-\tau')} \mathrm{d}\tau' + \eps f &= 0, \\
  \mathrm{i} d_2 \partial_\tau^2 b -(\eps+\mathrm{i}\zeta_0 )b + \mathrm{i}\big(|b|^2 + \frac{2}{2\ell}\int_{-\ell}^{\ell} |a(\tau')|^2 \mathrm{d}\tau' \big) b
  + \mathrm{i} \frac{1}{2\ell} \sum_\mu \beta_\mu\int_{-\ell}^{\ell} a(\tau') \mathrm{e}^{\mathrm{i}\mu \frac{\pi}{\ell} (\tau-\tau')} \mathrm{d}\tau' &= 0,
\end{split}
\end{align}
where we introduced the resonator length $2\ell>0$ as an additional parameter and use $\eps \geq 0$ to scale the effect of damping and forcing. In particular, both $a$ and $b$ are assumed to be $2\ell$-periodic in $\tau$. Note that~\eqref{eq:LLE_Stationary} coincides with the stationary version of (2) in the main text when $\ell = \pi$ and $\eps = 1$. To derive the soliton existence boundaries, we proceed in two steps. First, we derive a limit system for~\eqref{eq:LLE_Stationary} as $\ell \to \infty$. In this limit, the nonlocal terms in~\eqref{eq:LLE_Stationary} reduce to local coupling terms, which are simpler to analyze. Second, we study the limit equation and construct FWD and BWD solitons perturbing from the $\eps = 0$ case. Persistence conditions in the perturbation argument then yield the soliton existence boundaries. 

\subsection{Large cavity limit and dissipation as a perturbation} 
Here, we consider the coupled equations in the large cavity limit $\ell \to \infty$. Practically, this is equivalent to assuming that the pulse is short, with a spectral bandwidth that exceeds by far the bandwidth of the dispersion modification around the pump laser.
For the FWD and BWD waves, we use the ansatz of a localized solution supported on a CW background,
\begin{align}\label{eq:soliton_ansatz}
    a(\tau) = a_\textup{CW} + a_1(\tau), \qquad
    b(\tau) = b_\textup{CW} + b_1(\tau),
\end{align}
where $a_\textup{CW},b_\textup{CW} \in \mathbb{C}$, and $a_1,b_1$ are exponentially localized waveform profiles. Although such a solution ansatz does not yield a true periodic solution (unless $a_1(\tau) = b_1(\tau) = 0$), it gives an excellent approximation provided that the resonator length $\ell$ is large or the profiles sufficiently localized. Exponential localization of the solution profiles yields that the integrals $\int_{-\infty}^\infty |a_1|^j ,\int_{-\infty}^\infty |b_1|^j$ are finite for $j=1,2$. Thus inserting the ansatz~\eqref{eq:soliton_ansatz} into~\eqref{eq:LLE_Stationary}, noting that all integrals over the localized profiles are of order $O(\ell^{-1})$, and taking the limit $\ell \to \infty$, we obtain the limit problem
\begin{align}\label{eq:limit}
\begin{split}
  \mathrm{i} d_2 \partial_\tau^2 a -(\eps+\mathrm{i}\zeta_0 )a + \mathrm{i}\big(|a|^2 + 2 |b_\textup{CW}|^2 \big) a
  + \mathrm{i} \beta_0 b_\textup{CW} + \eps f &= 0, \\
  \mathrm{i} d_2 \partial_\tau^2 b -(\eps+\mathrm{i}\zeta_0 ) b + \mathrm{i}\big(|b|^2 + 2 |a_\textup{CW}|^2 \big)  b
  + \mathrm{i} \beta_0 a_\textup{CW} &= 0.
\end{split}
\end{align}
Since no exact soliton solutions are known for the coupled system~\eqref{eq:limit}, we construct DKS perturbatively for $0<\eps\ll1$. 
We note that $0<\eps\ll1$ is equivalent to the situation with full damping and driving ($\eps=1$), but where the detuning and pump power are large. Specifically, we can translate the results of the analysis presented below to the case of full dissipation in a finite size cavity via the scaling laws
$\tilde{a}(\tau)=\eps^{-1/2}a(\eps^{-1/2}\tau)$, $\tilde{b}(\tau)=\eps^{-1/2}b(\eps^{-1/2}\tau)$. The rescaled functions are then defined on the interval $[-\ell\eps^{1/2},\ell\eps^{1/2}]$ and solve~\eqref{eq:LLE_Stationary} with $\zeta_0,f,\beta_\mu,\ell$ replaced by
$ \tilde\zeta_0=\eps^{-1}\zeta_0,  \tilde\beta_\mu=\eps^{-1}\beta_\mu, \tilde f=\eps^{-1/2}f, \tilde\ell=\ell\eps^{1/2}$, and with the damping normalized to $1$. Note that $\tilde\ell=\pi$ can always be ensured by adjusting $d_2$, if necessary. With this choice, the rescaled system coincides with the stationary version of (2) in the main text. Thus, our analysis shows that the soliton approximation formulas and existence conditions are asymptotically sharp in the limit of large detuning and pump strength, and for rescaled coupling coefficients $\tilde\beta_\mu$.

\subsection{Analysis of the limit system~\eqref{eq:limit}}
%We construct DKS solutions for~\eqref{eq:limit} in the small damping and small forcing regime $0< \eps \ll 1$. 
Let us start with the CW background on which the DKS are formed. Expanding the background as $ a_\textup{CW}= \eps a_\textup{CW}^{(1)} + O(\eps^2)$, $ b_\textup{CW}= \eps b_\textup{CW}^{(1)} + O(\eps^2)$ and inserting these expansions into~\eqref{eq:limit} yields
\begin{align*}
    &\begin{pmatrix}
        -\mathrm{i} \zeta_0 & \mathrm{i} \beta_0 \\
        \mathrm{i} \beta_0 & -\mathrm{i} \zeta_0
    \end{pmatrix}
    \begin{pmatrix}
        a_\textup{CW}^{(1)} \\ b_\textup{CW}^{(1)}
    \end{pmatrix}
    =
    \begin{pmatrix}
        -f \\ 0
    \end{pmatrix}
    \quad
    \implies
    \quad
    \begin{pmatrix}
        a_\textup{CW}^{(1)} \\ b_\textup{CW}^{(1)}
    \end{pmatrix}
    =
    -\frac{\mathrm{i}f}{\zeta_0^2 - \beta_0^2}
    \begin{pmatrix}
        \zeta_0 \\ \beta_0
    \end{pmatrix},
\end{align*}
provided that $\zeta_0^2 \neq \beta_0^2$. Thus, we find a CW solution given by
\begin{align*}
    a_\textup{CW}= -\eps \frac{\mathrm{i}f\zeta_0}{\zeta_0^2-\beta_0^2} + O(\eps^2),\quad
    b_\textup{CW}= -\eps \frac{\mathrm{i}f\beta_0}{\zeta_0^2-\beta_0^2} + O(\eps^2).   
\end{align*}
Using the expansions of the backgrounds we find the equations for the soliton profiles
\begin{align}\label{eq:FWD_limit}
\begin{split}
  &-d_2 \partial_\tau^2 a_1 +\zeta_0 a_1 - |a_1|^2 a_1  
   + \eps \mathrm{i}\big( -  a_1 + \frac{2f \zeta_0}{\zeta_0^2 -\beta_0^2}|a_1|^2  - \frac{f \zeta_0}{\zeta_0^2 - \beta_0^2} a_1^2 \big) + O(\eps^2) =0
\end{split}
\end{align}
and
\begin{align}\label{eq:BWD_limit}
\begin{split}
  &-d_2 \partial_\tau^2 b_1 +\zeta_0 b_1 - |b_1|^2 b_1 
  + \eps\mathrm{i} \big( -  b_1 + \frac{2f \beta_0}{\zeta_0^2 -\beta_0^2}|b_1|^2  - \frac{f \beta_0}{\zeta_0^2 - \beta_0^2} b_1^2\big)  + O(\eps^2) =0.
\end{split}
\end{align}
Note that up to the order $O(\eps)$, equations~\eqref{eq:FWD_limit} and \eqref{eq:BWD_limit} decouple and hence we can analyze them independently.

\subsubsection{Analysis of FWD solitons}
We first construct a FWD DKS by solving \eqref{eq:FWD_limit} with a localized pulse solution while keeping $b_1 \equiv 0$, so that the BWD field remains in a CW state. To this end we insert $a_1 = a^{(0)} + \eps a^{(1)} + O(\eps^2)$ into~\eqref{eq:FWD_limit}, and obtain at order $O(\eps^0)$ and $O(\eps)$, respectively,
\begin{align*}
  -d_2 \partial_\tau^2 a^{(0)} +\zeta_0 a^{(0)} - |a^{(0)}|^2 a^{(0)} = 0&, \\
  L a^{(1)} = \mathrm{i}\big( a^{(0)} - \frac{2f \zeta_0}{\zeta_0^2 -\beta_0^2}|a^{(0)}|^2  + \frac{f \zeta_0}{\zeta_0^2 - \beta_0^2} (a^{(0)})^2 \big).&
\end{align*}
Here, the linear operator $L$ is given by 
$$
    L a^{(1)} = -d_2 \partial_\tau^2 a^{(1)} + \zeta_0 a^{(1)} - 2 |a^{(0)}|^2 a^{(1)} - (a^{(0)})^2 (a^{(1)})^*.
$$
At leading order we recover the nonlinear Schrödinger equation, which we solve using the bright-soliton solution
$$
    a^{(0)}(\tau) = \alpha_1 \mathrm{sech}(\alpha_2 \tau) \mathrm{e}^{\mathrm{i} \theta}, \quad
    \alpha_1 = \sqrt{2\zeta_0} ,\quad \alpha_2 =\sqrt{\zeta_0/d_2}.
$$
The phase $\theta$ of the soliton is determined by the solvability condition for the linear inhomogeneous equation for $a^{(1)}$. By the Fredholm alternative, this equation admits a solution if and only if the integral
\begin{align*}
	\textup{Re} \int_{-\infty}^\infty \big( a^{(0)} - \frac{2f \zeta_0}{\zeta_0^2 -\beta_0^2}|a^{(0)}|^2  - \frac{f \zeta_0}{\zeta_0^2 - \beta_0^2} (a^{(0)})^2 \big)(a^{(0)})^* \mathrm{d} \tau'
\end{align*}
vanishes. Evaluating this integral yields the phase equation
\begin{align}\label{eq:FWD_phase_condition}
	\frac{f \zeta_0^2 \pi}{\zeta_0^2 - \beta_0^2} \cos(\theta) = 2 \sqrt{2\zeta_0},
\end{align}
which can be solved provided that the parameters satisfy
\begin{align}\label{eq:FWD_condition_derivation}
	\frac{f \zeta_0^2 \pi}{|\zeta_0^2 - \beta_0^2|} \geq  2 \sqrt{2\zeta_0}.
\end{align}
This is the FWD DKS existence condition (equivalent to inequality (3) in the main text). 
As a consequence, if~\eqref{eq:FWD_condition_derivation} holds true, we find a FWD DKS that obeys the approximation formula
\begin{empheq}[box=\fbox]{align}
\begin{split}
	a(\tau) &\approx \sqrt{2\zeta_0} \mathrm{sech}(\sqrt{\zeta_0/d_2}\tau) \mathrm{e}^{\mathrm{i} \theta} - \eps \frac{\mathrm{i}f\zeta_0}{\zeta_0^2-\beta_0^2}, \\
	b(\tau) &\approx -\eps \frac{\mathrm{i}f\beta_0}{\zeta_0^2-\beta_0^2},
\end{split}
\end{empheq}
where $\theta$ is a solution to the phase equation~\eqref{eq:FWD_phase_condition} and $\eps$ is sufficiently small. We recall that this formula is derived in the long cavity limit $\ell \to \infty$, and thus provides an accurate approximation for large resonator lengths.

\subsubsection{Analysis of BWD solitons}
Next, we proceed with the construction of BWD DKS. Following the same arguments as for the FWD solitons, we first derive from ~\eqref{eq:BWD_limit} the BWD phase equation
\begin{align}\label{eq:BWD_phase_condition}
	\frac{f \zeta_0 \beta_0 \pi}{\zeta_0^2 - \beta_0^2} \cos(\theta) = 2 \sqrt{2\zeta_0},
\end{align}
for which a solution exists if the parameters satisfy
\begin{align}\label{eq:BWD_condition_derivation}
	\frac{f \zeta_0 |\beta_0| \pi}{|\zeta_0^2 - \beta_0^2|} \geq 2 \sqrt{2\zeta_0}.
\end{align}
This inequality is equivalent to the BWD DKS existence condition (4) in the main text. The approximation formula for BWD DKS then reads
\begin{empheq}[box=\fbox]{align}
\begin{split}
	a(\tau) &\approx - \eps \frac{\mathrm{i}f\zeta_0}{\zeta_0^2-\beta_0^2}, \\
	b(\tau) &\approx \sqrt{2\zeta_0} \mathrm{sech}(\sqrt{\zeta_0/d_2}\tau) \mathrm{e}^{\mathrm{i} \theta} -\eps \frac{\mathrm{i}f\beta_0}{\zeta_0^2-\beta_0^2},
\end{split}
\end{empheq}
where $\theta$ solves the BWD phase equation~\eqref{eq:BWD_phase_condition}.

\subsubsection{Analysis of simultaneous FWD-BWD solitons}
If both existence conditions~\eqref{eq:FWD_condition_derivation},~\eqref{eq:BWD_condition_derivation} are satisfied at the same time, then our analysis predicts the existence of simultaneously propagating FWD and BWD DKS of the form
\begin{empheq}[box=\fbox]{align}
\begin{split}
	a(\tau) &\approx \sqrt{2\zeta_0} \mathrm{sech}(\sqrt{\zeta_0/d_2}\tau) \mathrm{e}^{\mathrm{i} \theta_1} - \eps \frac{\mathrm{i}f\zeta_0}{\zeta_0^2-\beta_0^2}, \\
	b(\tau) &\approx \sqrt{2\zeta_0} \mathrm{sech}(\sqrt{\zeta_0/d_2}\tau) \mathrm{e}^{\mathrm{i} \theta_2} -\eps \frac{\mathrm{i}f\beta_0}{\zeta_0^2-\beta_0^2},
\end{split}
\end{empheq}
where $\theta_1,\theta_2$ solve~\eqref{eq:FWD_phase_condition},~\eqref{eq:BWD_phase_condition}, respectively.

\printbibliography

% \end{document}

\end{refsection}

\end{document}